\documentclass[
 preprint,
 superscriptaddress,
 amsmath,amssymb,
 aps,
 prl,
 longbibliography,
 showpacs
]{revtex4-1}
\usepackage{graphicx}
\usepackage{dcolumn}
\usepackage{bm}
\usepackage{epstopdf}
\usepackage{color}
\begin{document}


\title{Topological cavity based on slow light topological edge mode for broadband Purcell enhancement}

\author{Xin Xie}
\author{Sai Yan}
\author{Jianchen Dang}
\author{Jingnan Yang}
\author{Shan Xiao}
\affiliation{Beijing National Laboratory for Condensed Matter Physics, Institute of Physics, Chinese Academy of Sciences, Beijing 100190, China}
\affiliation{CAS Center for Excellence in Topological Quantum Computation and School of Physical Sciences, University of Chinese Academy of Sciences, Beijing 100049, China}
\author{Yunuan Wang}
\affiliation{Beijing National Laboratory for Condensed Matter Physics, Institute of Physics, Chinese Academy of Sciences, Beijing 100190, China}
\affiliation{Key Laboratory of Luminescence and Optical Information, Ministry of Education, Beijing Jiaotong University, Beijing 100044, China}
\author{Shushu Shi}
\author{Longlong Yang}
\author{Danjie Dai}
\author{Yu Yuan}
\author{Nan Luo}
\author{Ting Cui}
\author{Gaohong Chi}
\author{Zhanchun Zuo}
\email{zczuo@iphy.ac.cn}
\affiliation{Beijing National Laboratory for Condensed Matter Physics, Institute of Physics, Chinese Academy of Sciences, Beijing 100190, China}
\affiliation{CAS Center for Excellence in Topological Quantum Computation and School of Physical Sciences, University of Chinese Academy of Sciences, Beijing 100049, China}
\author{Bei-Bei Li}
\affiliation{Beijing National Laboratory for Condensed Matter Physics, Institute of Physics, Chinese Academy of Sciences, Beijing 100190, China}
\affiliation{Songshan Lake Materials Laboratory, Dongguan, Guangdong 523808, China}
\author{Can Wang}
\affiliation{Beijing National Laboratory for Condensed Matter Physics, Institute of Physics, Chinese Academy of Sciences, Beijing 100190, China}
\affiliation{CAS Center for Excellence in Topological Quantum Computation and School of Physical Sciences, University of Chinese Academy of Sciences, Beijing 100049, China}
\affiliation{Songshan Lake Materials Laboratory, Dongguan, Guangdong 523808, China}
\author{Xiulai Xu}
\email{xlxu@iphy.ac.cn}
\affiliation{Beijing National Laboratory for Condensed Matter Physics, Institute of Physics, Chinese Academy of Sciences, Beijing 100190, China}
\affiliation{CAS Center for Excellence in Topological Quantum Computation and School of Physical Sciences, University of Chinese Academy of Sciences, Beijing 100049, China}
\affiliation{Songshan Lake Materials Laboratory, Dongguan, Guangdong 523808, China}

\date{\today}

\begin{abstract}
Slow light in topological valley photonic crystal structures offers new possibilities to enhance light-matter interaction. We report a topological cavity based on slow light topological edge mode for broadband Purcell enhancement. The topological edge modes with large group indices over 100 can be realized with a bearded interface between two topologically distinct valley photonic crystals, featuring the greatly enhanced Purcell factor because of the increased local density of states. In the slow light regime, the topological cavity supports much more cavity modes with higher quality factor than that in the fast light regime, which is both demonstrated theoretically and experimentally. We demonstrate the cavity enables the broadband Purcell enhancement together with substantial Purcell factor, benefiting from dense cavity modes with high quality factor in a wide spectral range. It has great benefit to the realization of high-efficiency quantum-dot-based single-photon sources and entangled-photon sources with less restriction on spectral match. Such topological cavity could serve as a significant building block toward the development of photonic integrated circuits with embedded quantum emitters.
\end{abstract}
\maketitle

\section{\label{sec1}Introduction}
Slow light with a remarkably low group velocity is a fascinating physical effect, which outperforms fast light in many aspects \cite{baba2008nphotslow,krauss2008we,thevenaz2008slow}. For example, it can be used for buffering and time-domain processing of optical signals \cite{tucker2005slow,okawachi2005tunable,beggs2008ultracompact}. Furthermore, slow light can also promote stronger light-matter interaction, offering new possibilities for miniaturization and improvement of photonic devices \cite{rao2007singleprb,rao2007single,lund2008experimental,monat2009slow,ek2014slow,liu2014random,colman2010temporal}. Photonic crystal structures are particularly attractive for generating slow light due to the great potential applications in photonic integrated circuits. In photonic crystal waveguide, slow light has been demonstrated in the vicinity of the photonic band edge \cite{notomi2001extremely,letartre2001group}, which can be used to improve lasing characteristics \cite{ek2014slow,liu2014random} and realize efficient on-chip single-photon gun \cite{rao2007singleprb,rao2007single}. However, in the slow light regime, the extrinsic scattering loss has a profound impact on the propagation of light, limiting the further application of slow light devices.

Topological photonics provide a robust way to manipulate light. Especially, the topological edge state occurred at the interface between two topologically distinct structures exhibits robust unidirectional transport against defects and perturbations, which can be used as one-way waveguide \cite{haldane2008possible,wang2009observation,fang2012realizing,hafezi2011robust,khanikaev2013photonic}. Moreover, the topological resonators can be formed due to the robustness against sharp bends, which can be used to realize topological lasers \cite{bahari2017nonreciprocal,harari2018topological,bandres2018topological,yang2020spin,zeng2020electrically,gong2020topological,noh2020experimental} and chiral quantum optical interface \cite{barik2020chiral,mehrabad2020chiral}. Introducing slow light into topological structures offers a promising solution for taking the advantage of slow light and also topological protection, such as the enhancement of light-matter interaction and robustness against extrinsic scattering loss. So far, a few novel ideas have been proposed to realize slow light topological waveguides in different structures, such as gyromagnetic photonic crystal \cite{yang2013experimental,chen2019strong}, bianisotropic metamaterials \cite{chen2015manipulating} and topological valley photonic crystal (VPC) \cite{yoshimi2020slow,arregui2021quantifying}. In contrast to other structures, topological VPC is all-dielectric structure, which can be easily implemented in optical regime \cite{he2019silicon,shalaev2019robust}. Furthermore, it possesses good compatibility with embedded quantum emitters such as quantum dots (QDs), which can be used to realize chiral quantum interface \cite{barik2020chiral,mehrabad2020chiral}. In VPC, the slow light topological valley edge state can be realized at a bearded interface, which exhibits high group index as well as robust transport \cite{yoshimi2020slow,arregui2021quantifying}. The slow light mode provides a new platform to design topological cavity for further enhancing the light-matter interaction and exploring novel phenomena.

Here, we propose a topological cavity based on slow light topological edge modes for broadband Purcell enhancement. The topological cavity is formed using super-triangle with a bearded interface between two topologically distinct VPCs. Topological edge modes with large group indices over 100 can be realized at the bearded interface. The Purcell factor in an infinite topological waveguide is enhanced with the increase of group index. In the slow light regime, the topological cavity supports much denser cavity modes with high quality factor (Q) than that in the fast light regime, which is demonstrated both theoretically and experimentally. In such a topological cavity, we realize broadband Purcell enhancement with substantial Purcell factor of single quantum emitter, due to the existence of dense cavity modes with high Q in a wide spectral range. It relaxes the demand on spectral match between cavity modes and quantum emitters, having great potential in the development of highly efficient on-chip single-photon sources and entangled-photon sources. This work paves a way for exploring novel topological slow light devices with diverse functionalities in integrated photonic circuit platform.

\section{\label{sec2}Topological slow light edge mode}
\begin{figure}
\centering
\includegraphics[scale=0.43]{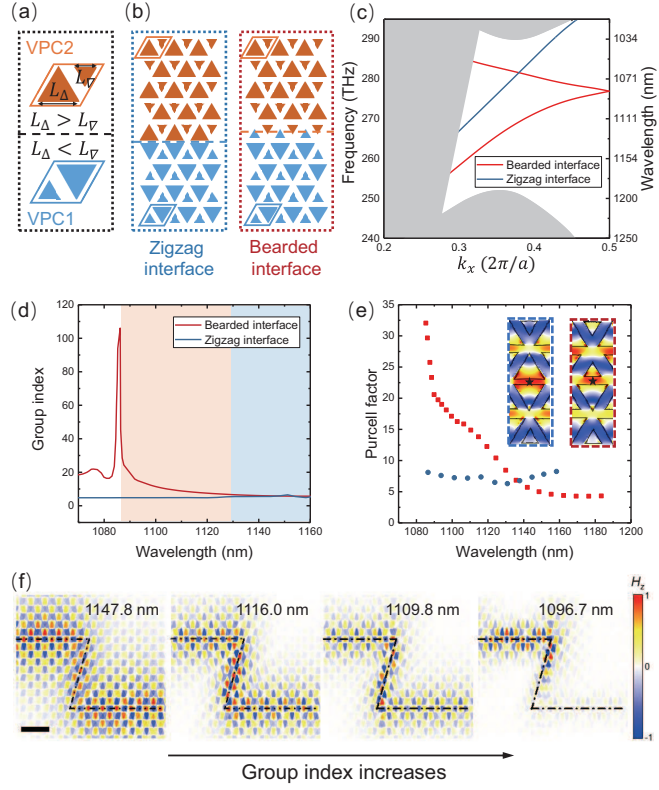}
\caption{ (a) Schematics of rhombic units of two topologically distinct VPCs, VPC1 (blue) and VPC2 (orange). They have different orientations of the larger triangle. In VPC1, $L_\bigtriangleup$ is smaller than $L_\bigtriangledown$, while in VPC2, $L_\bigtriangleup$ is larger than $L_\bigtriangledown$. (b) Schematics of the 2D topological VPC waveguide with zigzag (left) and bearded interface (right). The blue and orange regions represent two topologically distinct VPCs, VPC1 and VPC2. (c) Dispersion curves for the edge states formed at zigzag (blue line) and bearded (red lines) interface with $a=$ 340 nm, $L_l=253$ and $L_s=$ 138 nm, where $L_l$ ($L_s$) is the size of larger (smaller) triangular airholes. The edge states at the bearded interface have the slow light regime near the BZ edge. (d) Calculated group indices of the edge states as a function of wavelength at the zigzag (blue line) and bearded (red line) interface. The blue and orange regions correspond to the fast light and slow light regime at the bearded interface, respectively. (e) Calculated Purcell factor of $x$-polarized dipole source in an infinite waveguide with zigzag (blue dots) and bearded (red dots) interface as a function of wavelength. Insets show examples of electric field profile of $x$ component ($|E_x|^2$) of edge modes at the zigzag (left) and bearded (right) interface. The $x$-polarized dipole source was placed at the antinode of corresponding edge modes, as illustrated by the black stars in the insets. (f) Magnetic field distribution ($H_z$) of several edge modes propagating through a Z-shaped bearded interface. The boundary is indicated by the black dashed lines. The scale bar is 2 $\mu$m. From left to right, the wavelengths of edge states decrease and the group indices increase. These simulations were performed by FDTD method.}
\label{f1}
\end{figure}

The topological VPC investigated is composed of a honeycomb lattice with two inverted equilateral triangular airholes (pointing-up and pointing-down). Different sizes between the two triangular airholes break the spatial inversion symmetry, resulting in the formation of a band gap. The VPC features non-zero Berry curvatures, with opposite signs at K and K’ valleys. Although the Berry curvature integrated over the entire Brillouin zone (BZ) is zero due to the time-reversal symmetry, the valley Chern number $C_{K/K'}$, an integration of Berry curvature over the half of the BZ around K/K’ points, gives rise to a non-zero value \cite{he2019silicon,shalaev2019robust}. Figure \ref{f1}(a) shows the rhombic units of two topologically distinct VPCs, VPC1 (blue) and VPC2 (orange), with different orientations of the larger triangular airholes. By interchanging the sizes of two triangular airholes, the sign of the Berry curvature at each valley will be flipped, leading to opposite valley Chern numbers for VPC1 and VPC2. At the interface between the two topologically distinct VPCs, topological edge states are formed. Detailed discussions are shown in Supplementary Information.

Figure \ref{f1}(b) shows two types of interfaces, in which the left panel shows a zigzag interface, and the right panel shows a bearded interface. Both interfaces are formed between two topologically distinct VPCs, VPC1 (upper half) and VPC2 (lower half),  supporting topological edge states. The zigzag interface faces larger or smaller airholes at the boundary. This type of interface has been predominantly investigated for topological waveguides \cite{he2019silicon,shalaev2019robust} and cavities \cite{barik2020chiral,noh2020experimental,gong2020topological}. It supports fast light modes by crossing the bandgap with near-linear dispersion (shown as blue line in Fig. \ref{f1}(c)). In contrast, the bearded interface joints smaller airholes. The decrease of airhole size adjacent to interface leads to the local increase of refractive index. As a result, the frequency of topological edge mode decreases, and a trivial mode near the upper bulk band edge emerges (shown as red lines in Fig. \ref{f1}(c)). The two modes are degenerate at the BZ edge due to glide plane symmetry of the interface. Therefore, a heavily dispersive characteristics within the bandgap is realized at such an interface, leading to the formation of topological slow light mode. The dispersion curves were calculated for transverse electric field mode by the three-dimensional finite-difference time-domain (FDTD) method. The simulation parameters for the structure were set as $a=340$ nm, $L_l=253$ nm, $L_s=138$ nm, $n$ = 3.43, where $a$ is the period of unit cell, $n$ is the refractive index and $L_l$ ($L_s$) is the size of larger (smaller) triangular airholes. Compared to circular airholes \cite{mehrabad2020chiral}, the design using triangular airholes supports a single-mode topological slow light, as well as a direct and larger bandgap at K (K’) point for the same degree of asymmetry.

Figure \ref{f1}(d) shows the calculated group indices of the edge states at the bearded (red) and zigzag (blue) interface. Compared to zigzag interface, the bearded interface exhibits both fast light and slow light topological modes, as shown blue and orange regions in Fig. \ref{f1}(d) respectively. At the bearded interface, the group index shows a sharp increase near the BZ edge and reaches its highest value of about 106 at the wavelength of 1086.23 nm. This value is about 22 times higher than that of the topological edge mode at the zigzag interface. The increase of group index leads to an increase of local density of states (LDOS) in waveguide, therefore resulting in an enhancement of light-matter interaction. In particular, the spontaneous emission rate can be greatly enhanced as the Purcell factor scales with the group index \cite{rao2007singleprb}. Therefore, the topological slow light mode can support much larger Purcell factor than the fast light mode. Figure \ref{f1}(e) shows the calculated Purcell factor in an infinite waveguide with zigzag (blue) and bearded (red) interface as a function of wavelength. The simulations were performed by placing the $x$-polarized dipole at antinode of waveguide mode, as shown in the insets. As expected, the Purcell factor is barely changed with wavelength at the zigzag interface. In contrast, a significant enhancement in Purcell factor is observed at the bearded interface, demonstrating the great potential of the topological slow light mode in enhancing light-matter interaction.

The topological edge states at the zigzag interface have been demonstrated to be robust against sharp bends of $60^{\circ}$ and $120^{\circ}$ in previous reports \cite{chen2017valley,he2019silicon,shalaev2019robust}. To visualize the transport property of the topological edge states at bearded interface, we calculated the field distribution of topological edge state in the Z-shaped waveguide with different wavelengths, as shown in Fig. \ref{f1}(f). The edge states were excited by an $x$-polarized dipole source. In the fast light regime, the field distribution does not show any beating in field intensity, confirming the robust light transmission. In the slow light regime, the modes with relatively small group index still exhibit robust transport. Therefore, the bearded interface supports topological slow light edge modes exhibiting robustness against sharp bends. However, with the decrease of wavelength (from left to right in Fig. \ref{f1}(f)), the group index increases, the reflection from sharp corner increases, leading to the formation of standing waves in the waveguide.
\section{\label{sec3} Topological cavity}

\begin{figure}[b]
\centering
\includegraphics[scale=0.43]{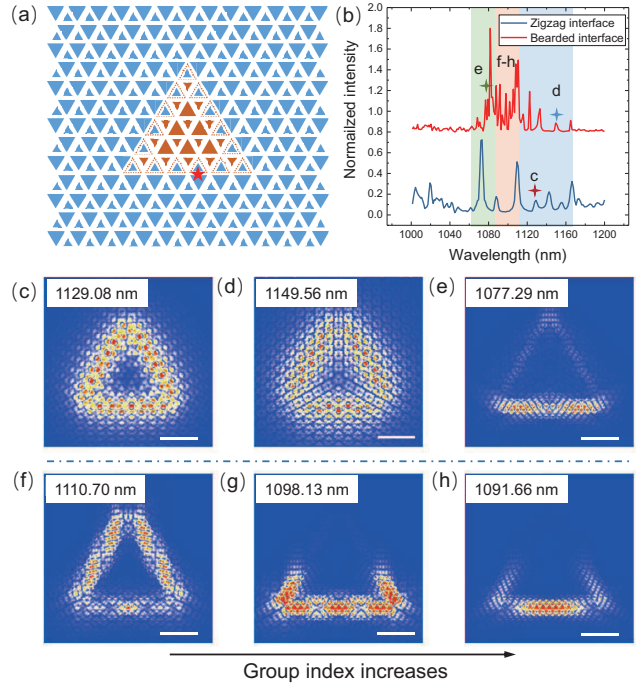}
\caption{ (a) Schematic of topological cavity shaped in the form of a super-triangle. (b) Simulated modes of the topological cavity with zigzag (blue line) and bearded (red line) interface. The spectra are shifted for clarity. (c) An example of electric field profile ($|E|$) of cavity mode in topological cavity with zigzag interface as indicated by the red star in (b). Examples of electric field profile of (d) topological modes with small group index, (e) trivial modes and (f-h) topological modes with high group index in topological cavity with bearded interface, corresponding to the blue, green and orange regions in (b). The modes in (d)(e) are indicated by the blue and green stars in (b). From (f) to (h), the wavelength decreases and the group index increase. The simulations were performed by putting a dipole source at the position indicated by the red star in (a). The scale bars in (c-h) are 2 $\mu$m.}
\label{f2}
\end{figure}
At the bearded interface, the topological edge modes still exhibit robust transport against sharp bends, except the modes with extremely high group index. This robustness enables them to form whispering gallery mode using a super-triangle. Figure \ref{f2}(a) shows the schematic of the topological cavity using a super-triangle. The inside (VPC2) and outside (VPC1) of the super-triangle correspond to two topologically distinct VPCs. Figure \ref{f2}(b) shows the calculated spectra from super-triangle cavity with zigzag (blue) and bearded (red) interface. The perimeters of the super-triangle cavity for zigzag and bearded interface are 44 unit cells (15 $\mu$m) and 50 unit cells (17 $\mu$m), respectively. With the zigzag interface, only the fast light mode can be used to form cavity mode, leading to a large free spectral range (FSR) about a dozen nanometers. Figure \ref{f2}(c) shows an example of electric field profile of cavity mode in the cavity with zigzag interface. The electric field distribution exhibits strong confinement at the topological interface and extends over the super-triangle. The large FSR results in a narrow-band Purcell enhancement, leading to the strict restriction on the spectral alignment between cavity mode and quantum emitters. While, with the bearded interface, both fast light and slow light modes are supported, which can be used to form cavity modes. In the fast light regime, the cavity modes show a similar behavior, including the large FSR and similar electric field profile. As the wavelength goes down, entering the slow light regime, the cavity modes become denser due to the increase of group index.

Figure \ref{f2}(d-h) show the electric field profile of the cavity modes in the three regions as shown in Fig. \ref{f2}(b). The electric field was excited by the dipole source located at the position indicated by the red star in Fig. \ref{f2}(a). In the blue region (in Fig. \ref{f2}(b)), corresponding to fast light mode and the slow light mode with relatively small group index, the electric field extends over the whole interface of the super-triangle, as shown in Fig. 2(d), demonstrating the robustness against sharp corners. However, in orange region for the slow light modes with high group index, they show different standing-wave patterns, as illustrated in Fig. \ref{f2}(f-h). With increasing group index (from Fig. \ref{f2}(f) to (h)), the cavity modes tend to localize at one side of the super-triangle, which may result from the increase of reflection from the sharp corners \cite{yoshimi2020slow,gong2020topological}. As for the green region, corresponding to the trivial mode, they are denoted as Fabry-Perot mode due to the lack of robustness against sharp corners, as shown in Fig. \ref{f2}(e). Note that the distinction between trivial mode and topological mode mainly depends on the edge dispersion in Fig. \ref{f1}(c).

\begin{figure}[t]
\centering
\includegraphics[scale=0.43]{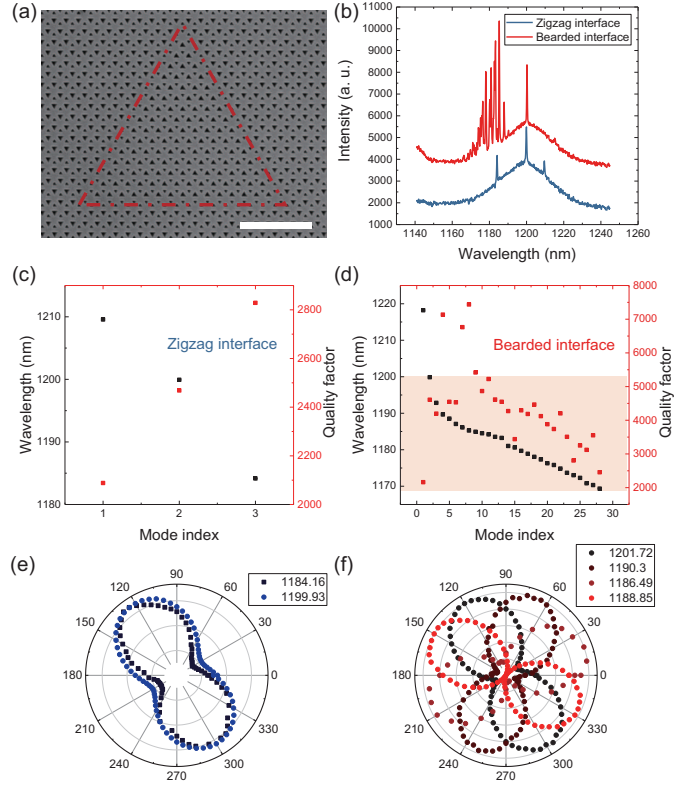}
\caption{ (a) SEM image of a fabricated topological cavity with a scale bar of 2 $\mu$m. The topological cavity is formed by super-triangle with bearded interface, as indicated by the red dashed line. (b) PL spectra from topological cavity with zigzag (blue line) and bearded (red line) interface. The spectra are shifted for clarity. (c-d) Wavelength (black squares) and corresponding Q (red squares) of the cavity modes in topological cavity with (c) Zigzag  and (d) bearded interface. The wavelength and Q were extracted by Lorentz-fitting of high-resolution spectra. In the orange region in (d), much denser cavity modes with high Q are supported in a wide spectral range about 30 nm. (e-f) Polar plots of linear-polarization dependence of several cavity modes in topological cavity with (e) zigzag and (f) bearded interface.    }
\label{f3}
\end{figure}
Then we fabricated the topological cavity into a GaAs slab with a thickness of 170 nm with an embedded layer of InAs QDs as light sources to probe the mode of the device. The structures were patterned using electron beam lithography followed by inductively coupled plasma etching. The sacrificial layer was removed with HF etching to form air bridge. Furthermore, the surface passivation treatment using Na$_2$S solution was performed to reduce the surface-state-related absorption losses and improve the Q \cite{kuruma2020surface}. Figure \ref{f3}(a) shows the scanning electron microscope (SEM) image of a fabricated cavity. The optical properties of the topological cavity were characterized by the confocal micro-photoluminescence (PL) measurement at low temperature (5 K) using a liquid helium flow cryostat.

Figure \ref{f3}(b) shows the PL spectra from cavities with zigzag (blue line) and bearded (red line) interface. In the cavity with a zigzag interface, three peaks are observed with an FSR about a dozen nanometers. The Qs are about 2000-3000, as shown in Fig. \ref{f3}(c). Whereas in the cavity with a bearded interface, much denser cavity modes with higher Q (up to 8000) are observed, especially in the wavelength range from 1169-1190 nm, corresponding to the slow light regime, as shown in Fig. \ref{f3}(d). The mode spectral spacing can be even below 1 nm. Higher Q in the slow light regime compared to those in the fast light regime is related to the increased LDOS together with topological protection. The Q reduces with the wavelength further decreasing, which may result from the topological protection becoming weaker, leading to an increased effect of scattering losses.

Additionally, we measured the linear-polarization dependence of these cavity modes with zigzag (Fig. \ref{f3}(e)) and bearded (Fig. \ref{f3}(f)) interface using half-wave plate (HWP) followed by a polarizer in the detection path. The cavity modes with zigzag interface have the same polarization direction. While in the cavity modes with bearded interface they are different, especially in the slow light regime. In fast light regime, the electric field distribution extending over the whole interface of super-triangle leads to a very close polarization property. In contrast, in slow light regime, the different polarizations may result from the different electric field distributions of the cavity modes excited by random QDs at different sides.

\section{\label{sec4}Discussion}

\begin{figure}[t]
\centering
\includegraphics[scale=0.43]{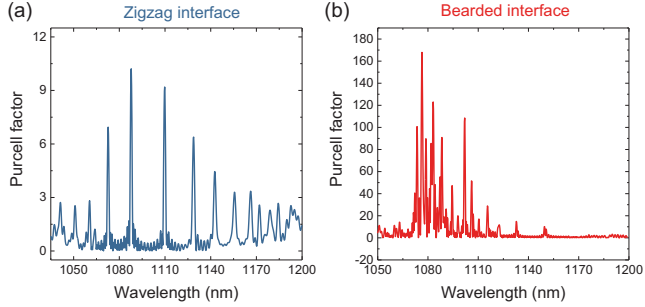}
\caption{ Calculated Purcell factor of topological cavity as function of wavelength with (a) zigzag and (b) bearded interface.  }
\label{f4}
\end{figure}

The cavity with bearded interface investigated here features much denser cavity modes with high Q in a broad spectral range. The dense cavity modes make it much easier to get resonance with quantum emitters located in a wide spectral range, enabling broadband Purcell enhancement. Besides broadband Purcell enhancement, the cavity modes in this regime also enable stronger light-matter interaction due to the increase of LDOS. Figure \ref{f4} shows the calculated Purcell factor of topological cavity as a function of wavelength with zigzag (Fig. \ref{f4}(a)) and bearded (Fig. \ref{f4}(b)) interface. As expected, the topological cavity with a zigzag interface exhibits a narrow-band Purcell enhancement with small Purcell factor about 10. In the fast light regime, the topological cavity with bearded interface shows a similar behavior to the topological cavity with zigzag interface. Whereas, in the slow light regime, it exhibits much stronger Purcell enhancement in a broad spectral range with large Purcell factor up to 170 due to the increased LDOS.

Compatibility with QDs, such a cavity has outstanding advantages in the development of single-photon sources and entangled-photon sources. It is well known that self-assembled QDs have been demonstrated to be a promising candidate to realize single-photon sources \cite{michler2000quantum,xu2007,senellart2017high} and entangled-photon sources \cite{benson2000regulated,stevenson2006semiconductor,akopian2006entangled}. By coupling to cavity, the spontaneous emission rate can be enhanced and the efficiency can be greatly improved \cite{purcell1946resonance}. To gain a strong enhancement, it is essential to get a good spectral and spatial match between cavity mode and quantum emitters, as well as a cavity mode with small mode volume and high Q. However, it is usually challenging due to the random distribution of QD both in position and emission wavelength, especially in the cavities with smaller mode volume or length with large FSR, exhibiting a narrow-band Purcell enhancement \cite{englund2005controlling,liu2018high,qian2019enhanced,xie2020cavity}. By lowing the Q, the broadband enhancement can be realized, which has been used to realize entangled-photon sources \cite{liu2019solid}. However, the lower Q reduces the Purcell factor. In contrast, for a large cavity, although the FSR is small and Q is high, the mode volume is large, leading to a weak light-matter interaction strength. Therefore, a cavity which can support broadband enhancement as well as large Purcell factor is necessary for the realization of high-efficient single-photon sources and especially, entangled-photon sources which need both enhancement of biexciton and exciton with different emission wavelengths.

In the topological cavity proposed here, the dense cavity modes in a wide spectral range make it much easier to tune quantum emitters into resonance with cavity mode. Thus the Purcell enhancement of quantum emitters located in the wide spectral range can be realized with less restriction on spectral match. Meanwhile, the dense cavity modes also enable simultaneous enhancement of biexciton and exciton with different emission wavelengths, which can be used to realize entangled-photon sources. Furthermore, the increased LDOS resulting from the slow light effect will give rise to strong Purcell enhancement. Therefore, this topological cavity enabling broadband Purcell enhancement with substantial Purcell factor provides an ideal platform to realize highly efficient single-photon sources and entangled-photon sources with less restriction on spectral match.

\section{\label{sec5} Conclusion and outlook}
In summary, we have proposed a topological cavity based on the topological slow light mode in VPC for broadband Purcell enhancement. Topological edge state with high group index over 100 can be realized at the bearded interface of VPC. Substantial enhancement in Purcell factor with the slow light is demonstrated. By comparing the topological cavity with zigzag and bearded interface both in theory and experiment, we found the slow light regime exhibits much denser cavity modes with high Q. Experimentally a Q up to 8000 is demonstrated with bearded interface. With numerical simulations, we have demonstrated that broadband Purcell enhancement with large Purcell factor can be realized in such a cavity.

This topological cavity, enabling broadband enhancement, provides a versatile platform to realize high-efficiency single-photon sources and entangled-photon sources. On the one hand, the broadband enhancement is beneficial to the realization of highly efficient single-photon sources with less restriction on spectral match. On the other hand, it can also be used to realize efficient entangled-photon sources, which needs both enhancement in biexciton and exciton with different emission wavelengths. Moreover, the existence of slow light mode in the cavity allows for the realization of photon-photon interaction due to the slow-down of light. By integrating with topological waveguides, a chiral quantum interface can be formed, forming the basis of scalable chiral quantum optical circuits. Exciting prospects can be predicted, including the development of complex nanophotonic circuits for quantum information processing and studying novel quantum many-body dynamics.

\begin{acknowledgments}
This work was supported by the National Natural Science Foundation of China (Grants No. 62025507, No. 11934019, No. 11721404, No. 61775232 and No. 11874419), the Key-Area Research and Development Program of Guangdong Province (Grant No.2018B030329001), and the Strategic Priority Research Program (Grant No. XDB28000000) of the Chinese Academy of Sciences.
\end{acknowledgments}

\end{document}